\documentclass[12pt]{article}
\usepackage{latexsym}

\def\tr{{\rm tr}}

\def\ket#1{\mid~\!\!\!{#1}~\!\!\rangle}
\def\bra#1{\langle~\!\!{#1}~\!\!\!\mid}
\def\cH{{\cal H}}

\def\IF{if and only if }
\def\QM{quantum mechanics }
\def\qm{quantum mechanics}
\def\QMl{quantum-mechanical }

\def\${\enskip$}
\def\M{measurement }
\def\m{measurement}
\def\Q{quantum }

\def\N{_{12\dots N}}

\begin{document}

{\bf \large \noindent Quantum
Correlations in Multipartite States.\\
Study Based on the Wootters-Mermin Theorem}\\

{\bf \noindent F. Herbut}\\

\vspace{0.5cm}

\rm

\noindent {\bf Abstract} \\
Decomposition of any \$N$-partite state
(density operator) into clusters (that
do not overlap) is studied in detail
with a view to learn as much as
possible about the correlations implied
by the state. The Wootters-Mermin
theorem, stating that the totality of
all strings of cluster events
(projectors) determines the state in
any finite- or infinite-dimensional
state space, is a slightly sharpened
and generalized form of the original
results of Wootters and Mermin. The
theorem is applied to tensor
factorization of the state into states
of clusters (uncorrelated
decomposition) and it is shown that a
finest uncorrelated decomposition
always exists, and that its coarsenings
and only they are other possible
uncorrelated cluster decompositions.
Distant effects within homogeneous
cluster states, which are, by
definition, the tensor factors in the
finest uncorrelated decomposition, are
discussed. The entire study is viewed
by the author as a possible further
elaboration of Mermin's Ithaca program.\\

Multipartite states. Coincidence of
subsystem events. Tensor factorization
of state.

\vspace{1.5cm}

{\bf \noindent 1. Introduction}\\

Mermin's "Ithaca interpretation" (I'd
rather call it the "Ithaca program")
\cite{Mermin1} was written in a very
inspiring way. It postulated that {\it
probability} must have a (yet unknown)
meaning for the individual quantum
system, and that it must be the basic
notion of \qm .

In the standard formalism of \QM the
basic concept is the state (a density
operator) of the system. We know from
Gleason's celebrated \noindent
{\footnotesize \rm \noindent
\rule[0mm]{4.62cm}{0.5mm}

\noindent F. Herbut (mail)\\
Serbian Academy of Sciences and Arts,
Knez Mihajlova 35, 11000 Belgrade,
Serbia\\
e-mail: fedorh@infosky.net and from USA
etc. fedorh@mi.sanu.ac.rs}\\

\noindent theorem \cite{Gleason} that
'state' and the entirety of
probabilities are equivalent. (See
perhaps the author's discussion of the
assumptions of Gleason's theorem in
\cite{FHZurek}, subsections 4.3 and
4.5.) Thus, the probabilities are the
{\it observable} aspect

\noindent of the state, and the latter
is, as Mermin puts it, an
'encapsulation' of the entirety of the
former.

In the case of {\it composite systems},
e. g., two particles far apart, only
subsystem (single-particle) observables
can be measured at best in coincidence.
First Wootters \cite{Wootters}, then
Mermin \cite{Mermin1} have called
attention to the important fact that
any composite-system state is
determined by averages of strings of
subsystem observables. Both authors
gave nice proofs of their claims.
Wootters' theory is critically
reproduced in Section 2. Mermin's
theorem is reproduced in Section 3 with
two slight elaborations: It is shown
that Mermin's subsystem observables can
be restricted to subsystem elementary
events (ray projectors), and that the
theorem is valid also in
infinite-dimensional (separable)
Hilbert spaces.

Naturally, one wants to know what kind
of states of composite systems one can
encounter and what the correlations can
do in them. Several researchers
\cite{Jordan}, \cite{Cabello59},
\cite{Cabello60}, \cite{Seevinck}
approached this problem in an inductive
way: by constructing concrete examples,
and then drawing conclusions from them.
On account of the results of Wootters,
of Mermin, and of these concrete
examples a correlations structure has
begun to show through a mist of
insufficient comprehension.

In this article the approach is
reversed: We derive a general theory of
composite-system states in a deductive
way leaning heavily on the
Wootters-Mermin theorem. Then we
discuss the mentioned examples of
composite-system states.

Seevinck \cite{Seevinck} seems to have
been carried away by his result. He
claims to have shown that Mermin's
Ithaca mantra "the correlations, not
the correlata" is disproved because, as
he maintains to have shown, the
correlations lack local reality.

'Reality' is a very serious question.
Some of us feel a kind of piety for
'reality'. I cannot put this better
than John S. Bell \cite{Bell}, though
his remark was aimed against the
instrumentalist approach, which shrinks
reality to correlating successive
instrumental readings:

\begin{quote}
"... experiment is a tool. The aim
remains: to understand the world. To
restrict \QM to be exclusively about
piddling laboratory operations is to
betray the great enterprise. A serious
formulation will not exclude the big
world outside the laboratory."
\end{quote}

Bell, a theorist, did not even mention
theoretical research efforts; his
priority was the experiment. But along
with his "piddling laboratory
operations" one can put 'piddling
probability calculations'.

I cannot imagine how can anybody doubt
the {\it reality} of {\it observable}
elements of nature. I would add two
requirements to the "Ithaca program"
\cite{Mermin1} to give 'reality' the
position that it deserves (though
Mermin might, perhaps, not agree):

a) If a \QMl entity (simple or complex)
is observable in the laboratory, then
it is real.

b) If the \QMl formalism establishes a
natural equivalence of two \QMl
entities (simple or complex) without
any arbitrariness, and if one of them
is real, then so is the other.

Requirement (b) is modeled  on the
example of Gleason's theorem
\cite{Gleason}, and reality of the
quantum state (density operator) is the
first and most fundamental application
of this requirement.\\

\vspace{0.5cm}

\noindent {\bf 2. The Theory of
Wootters}\\

A Hermitian \$M\times M\$ matrix can be
specified by giving the \$M\$ real
diagonal elements and the \$M(M-1)/2\$
complex elements  above the diagonal.
If the matrix has trace \$1\$, then it
contains \$(M-1)+M(M-1)=M^2-1\$
independent real numbers.

If one performs \$M^2-1\$ non-trivial
yes-no \m s, then the state of the
system is determined. We elaborate this
claim of Wootters \cite{Wootters} in
the Appendix.

Wootters came up with the fascinating
observation \cite{Wootters} that the
expounded \$(M^2-1)$-arithmetic result
{\it implies} that an \$n$-partite
composite system can be determined by
yes-no subsystem coincidence \m s, in
particular, by those that determine the
subsystem states. Let us see this in
more detail.

Let \$M_k\$ be the dimension of the
state space of the \$k$-th subsystem
(Wootters' theory was confined to
finite-dimensional state spaces). Then,
according to the above arithmetical
claim, one needs
\$\Big(\prod_{k=1}^nM_k\Big)^2-1\$
yes-no \m s, i. e., \m s of non-trivial
events (projectors) to specify the
composite-system state. To evaluate the
state of the \$k$-th subsystem, one
needs \$M_k^2-1\$ corresponding
non-trivial subsystem \m s. Consider
now strings of \$n\$ of these same
subsystem \m s including the certain
events (the identity operators) any
number of times. The number of these
strings can be obtained by joining the
certain event to each of the
\$(M^2_k-1)\$ yes-no \m s, making a set
of \$M_k^2\$ events, and by multiplying
these numbers: \$\prod_{k=1}^nM_k^2\$.
If we now subtract the coincidence of
certain events in all subsystems, i.
e., the certain event for the composite
system (it gives no information), we
obtain
\$\Big(\prod_{k=1}^nM_k\Big)^2-1\$, the
number of non-trivial yes-no \m s
needed to determine the
composite-system state  (density
matrix). (Note that the
composite-system event is non-trivial
if at least one of the subsystem events
in the string is non-trivial.)

Wootters himself comments on this
amazing result saying: "In this sense
\QM uses its information
economically."\\

\vspace{0.5cm}

\noindent {\bf 3. The Wootters-Mermin
Theorem}\\

\noindent As it was mentioned in the
Introduction, the fundamental
connection between probabilities of
{\it all} events (projectors) on the
one hand and states of quantum systems
(density operators) on the other is
established by the well-known theorem
of Gleason \cite{Gleason} (see also
\cite{FHZurek}, subsections 4.3 and
4.5). In case of composite systems
subsystems may be spatially apart from
each other, and two-subsystem or
more-subsystem events cannot be
realized in \m . Hence, if the
requirement of 'all events' in
Gleason's theorem could not be confined
to 'coincidences of subsystem events'
for composite systems, the state
concept, and all first-quantization \QM
would break down (into useless
fiction).

This is where the Wootters-Mermin
theorem (to be stated and proved) saves
\QM supplementing Gleason's theorem in
a satisfactory way.

The basic ingredients of the theorem
are the mentioned coincidences of
subsystem events apparently first
discovered by Wootters \cite{Wootters},
and aftewards independently discovered
by Mermin \cite{Mermin1}. But the
approach taken below in the formulation
of the theorem follows Mermin's ideas,
not those of Wootters, because the
latter's arithmetic does not allow
extension to infinity.\\

We assume that the state space is
finite or infinite dimensional. More
precisely, we have a complex, separable
Hilbert space. Further, we have in mind
an \$N$-partite, e. g., an
\$N$-particle \Q system. For
simplicity, we'll use the language of
particles.\\

{\bf Definition 1.} We envisage the set
\$\{1,2,\dots ,N\}\$ of all particles
in the system decomposed into \$n\$,
\$1\leq n\leq N\$, non-overlapping
classes the union of which gives the
entire set. We call the classes {\it
clusters}, and the decomposition a {\it
cluster decomposition} (CD). If
\$n=1\$, then the decomposition is said
to be trivial. If \$n=N\$, then the CD
is maximal.\\

Now a slightly sharpened and
generalized form of the results of
Wootters and Mermin, is presented as a
theorem. It is called the
Wootters-Mermin theorem, but also the
{\it CD theorem}.\\

{\bf Theorem 1.} Let us consider a {\it
composite system} in an arbitrary given
state \$\rho\$ and an arbitrary given
cluster decomposition (cf Definition
1). The state is {\it uniquely
determined} by the probabilities of
{\it all} subsystem coincidences
\$P_1\otimes P_2 \otimes \dots\otimes
P_n\$. Here each event \$P_k\$,
\$k=1,2,\dots ,n\$ (projector in
\$\cH_k\$, the state space of the
\$k$-th cluster) is an elementary event
(a ray projector) or the certain
event (the identity operator).\\

We will often write \$P_1P_2\dots P_n\$
instead of \$P_1\otimes P_2\otimes\dots
\otimes P_n\$ meaning by \$P_k\$
actually \$I_1\otimes\dots\otimes
P_k\otimes\dots\otimes I_n\$ for
\$1<k<n\$ in this context, where
\$I_k\$ is the identity operator in
\$\cH_k$. (For \$k=1\$ and \$k=n\$ the
required modification is obvious.)\\

{\bf Proof.} Let \$Q\$ be a projector
(event) in the state space of the given
composite system. Then, according to
the well-known theorem of Gleason
\cite{Gleason}, the totality of all
events \$Q\$ determines \$\rho\$
uniquely via the \QMl probability
formula
$$\tr(Q\rho ).\eqno{(1)}$$

We make restriction to finite-trace
projectors \$Q\$ because they, in
contrast to infinite-trace ones, do
belong to the Hilbert space of
Hilbert-Schmidt (HS) operators. This
makes the proof rigorous. Finite-trace
projectors are sufficient for the proof
because infinite-trace projectors are
expressible as limeses of finite-trace
ones, and probability is continuous.
Thus, the infinite-trace-projector
probabilities are implied by the
finite-trace-projector ones.

Let \$\{\ket{m_k}_k:m_k=1,2,\dots \}\$
be arbitrary complete orthonormal bases
in the state spaces \$\cH_k\$,
\$k=1,2,\dots ,n\$ such that the
quantum numbers are ordered (like the
natural numbers). Every mentioned
projector \$Q\$ can be represented in
this basis because
\$\prod_{i=1}^{\otimes N}\cH_i=
\prod_{k=1}^{\otimes n}\cH_k\$ (the
equality actually means isomorphism).
To write the representation in operator
form, one introduces the corresponding
dyads. Then the expansion reads
$$Q=\sum_{m_1}
\sum_{m'_1}\sum_{m_2}\sum_{m'_2}\dots
\sum_{m_n}\sum_{m'_n}
\bra{m_1}_1\bra{m_2}_2\dots \bra{m_n}_n
Q\ket{m'_1}_1\ket{m'_2}_2\dots
\ket{m'_n}_n\times$$
$$\ket{m_1}_1\bra{m'_1}_1
\ket{m_2}_2\bra{m'_2}_2\dots
\ket{m_n}_n\bra{m'_n}_n.\eqno{(2)}$$

To rid ourselves of the off-diagonal
dyads \$\ket{m_k}\bra{m'_k},\enskip
m_k<m'_k,\enskip k=1,2,\dots ,n\$, and
replace them by ray projectors, we
define
$$\forall (m_k<m_k'):\quad
\ket{m_k,m_k',1}\equiv (1/2)^{1/2}
(\ket{m_k}+\ket{m_k'}),$$
$$\ket{m_k,m_k',2} \equiv
(1/2)^{1/2}(\ket{m_k}-i\ket{m_k'}).$$
Inverting these definitions, one
obtains $$\forall (m_k<m_k'):\quad
\ket{m_k}\bra{m_k'}=
\ket{m_k,m_k',1}\bra{m_k,m_k',1}
-i\ket{m_k,m_k',2}\bra{m_k,m_k',2}+$$
$$[(i-1)/2] \ket{m_k}\bra{m_k}+
[(i-1)/2]
\ket{m_k'}\bra{m_k'}.\eqno{(3)}$$
Naturally, \$\ket{m'_k}\bra{m_k}=
(\ket{m_k}\bra{m'_k})^{\dag}$.

Next, we replace each off-diagonal dyad
by the linear combination of 4 ray
projectors according to (3) or its
adjoint. One should note that, while
(2), if a series, is absolutely
convergent, hence the order can be
changed by an arbitrary (infinite)
permutation. This is no longer true
when the expansion is exclusively in
diagonal dyads, which are ray
projectors. (The order still may be
permuted if the 4 dyads introduced by
(3) or its adjoint are kept together
for all \$k\not= k'\$.) The above
'diagonalization' procedure leading to
(3) is not unique.

All that remains to be done is to
substitute \$Q\$ in (1) by its
expansion (2) in which the off-diagonal
dyads have been replaced by ray
projectors according to (3) or its
adjoint. The sums (series), along with
the numbers can be taken outside the
trace due to the linearity (and
continuity) of the trace. Then we have
linear combinations (possibly infinite
ones) of probabilities of coincidences
of elementary cluster-events (ray
projectors). \hfill $\Box$\\

One should have in mind that the scalar
product in the Hilbert space of two HS
operators \$A\$ and \$B\$ is
\$(A,B)\equiv\tr\Big(
A^{\dag}B\Big)$.\\

The proof of Mermin's theorem presented
actually confines the strings of
subsystem ray projectors far more than
stated in the theorem. (They are all
generated from one fixed basis.) But we
will not utilize this stronger form.
Actually, we'll often make use of a
form that is even weaker than the
formulation of the theorem: we'll use
any subsystem events (not
just elementary ones).\\

\vspace{0.5cm}

\noindent {\bf 4. Immediate
Consequences\\}

\indent One may object that the
Wootters-Mermin theorem does not give a
practical way how to evaluate \$\rho\$
from the coincidence probabilities.
Neither does Gleason's theorem. I see
the former as an important elaboration
of Gleason's theorem in the case of
composite systems. The fundamental
significance
of both lies in their generality.\\

Next, one wonders what correlations
are. This is a very elusive concept.
The only case that I know when one can
put one's finger on an entity
expressing the correlations is the case
of bipartite state vectors, where the
(antiunitary) correlation operator
'carries' all correlations. (More about
this below, in Lemma 1.)

As it is in Gleason's case, where each
positive-probability event 'probes' the
state, in composite systems the strings
of subsystem events 'probe' the state,
and {\it ipso facto} they 'probe' the
correlations implied by the state. In
lack of a general definition of
correlations, there is a natural way
how to define sort of 'part' of the
correlations that a given string of
subsystem events 'sees' probing the
state.

In classical probability theory it is
well understood that events are
uncorrelated if the coincidence
probability equals the product of the
separate probabilities of the events.
Classical probability theory is
relevant because the projectors in a
string of subsystem events always
commute with each other. Hence, it
seems natural to make the following
definition {\it of the correlations
that a given string of subsystem events
'sees' probing the state}.\\

{\bf Definition 2.} Let \$\rho\$ be a
state of a composite system of \$N\$
particles, and let a CD be given that
breaks up the set of all particles into
\$n\$ clusters (cf Definition 1). Let,
further, \$P_1P_2\dots P_n\$ be a given
string of subsystem events - each
factor being a subsystem projector
possibly the identity operator. Then
the correlations that the string 'sees'
in \$\rho\$ is {\it the absolute value
of the difference between the
coincidence probability and the product
of subsystem-event probabilities}:
$$\Big|\tr\Big(\rho\prod_{k=1}^{\otimes n}
P_k\Big)-\prod_{k=1}^n\tr(\rho
P_k)\Big|.\eqno{(4)}$$  The subsystem
events are {\it uncorrelated} if the
string 'sees' correlations that are
zero; otherwise they are
{\it correlated}.\\

{\bf Corollary 1.} If the string in
Definition 2 contains a
zero-probability subsystem event, then
the string 'sees' no correlations.\\

{\bf Proof.} Let, e. g., \$\tr (\rho
P_1)=0\$. This implies \$P_1\rho
P_1=0\$ (because \$\tr(P_1\rho
P_1)=0\$, and a positive operator can
have zero trace only if it is zero
itself). Then \$\tr
\Big(\rho\prod_{k=1}^{\otimes
n}P_k\Big)= \tr \Big((P_1\rho P_1
)\prod_{k=1}^nP_k\Big)=0\$, and
analogously
\$\prod_{k=1}^n\Big(\tr(\rho
P_k)\Big)=0\$ (cf Definition 2).\hfill
$\Box$\\

Since zero-probability subsystem events
disable any string in which they appear
to 'see' correlations, it is best to
avoid them.

Positive-probability events need not
coincide, i.e., one can have, e. g.,
\$\tr\Big(\rho_1P_1\Big)>0
<\tr\Big(\rho_2P_2\Big)\$, and
\$\tr\Big(\rho_{12}(P_2\otimes
P_2)\Big)=0\$. Example:
\$P_1\equiv\ket{+}_1\bra{+}_1$, and
\$P_2\equiv\ket{+}_2\ket{+}_2\$ in the
singlet state.

If one does not take the absolute value
in (4), then a string of subsystem
events can 'see' {\it increase} or {\it
decrease} of coincidence probability
with respect to the product of
subsystem probabilities. The example
last mentioned involves decrease. The
example
\$P_1\equiv\ket{+}_1\bra{+}_1\$, and
\$P_2\equiv\ket{-}_2\bra{-}_2\$ in the
singlet state involves increase.\\

In the special case of a {\it bipartite
state vector} \$\ket{\Psi}_{12}\$ , the
(antiunitary) {\it correlation
operator} \$U_a\$, implied by the state
vector, is the carrier of the entire
correlations \cite{FHMV76},
\cite{FHZurek}. Hence, one can derive
the correlations (4) 'seen' by a given
string of subsystem events \$P_1,P_2\$.
This is done in the
next lemma.\\

{\bf Lemma 1.} Let \$\ket{\Psi}_{12}\$
be an arbitrary bipartite state vector,
and let \$P_1,P_2\$ be arbitrary
subsystem events (projectors). Then the
correlations 'seen' by these events are
$$\Big|\Big(\sum'_q\bra{q}_1\rho_1^{1/2}
U_a^{\dag}P_2U_a\rho_1^{1/2}\ket{q}_1\Big)
\enskip
-\enskip\Big(\tr(\rho_1P_1)\Big)
\tr(\rho_2P_2)\Big|,$$ where
\$\{\ket{q}_1:\forall q\}\$ is a
complete orthonormal basis in \$\cH_1\$
such that a subset of it spans the
range of \$P_1\$:
\$P_1=\sum'_q\ket{q}_1\bra{q}_1\$, the
prim on the sum denoting that one sums
only over the mentioned subset.
Naturally, \$\rho_i\equiv\tr_j\Big(
\ket{\Psi}_{12}\bra{\Psi}_{12}\Big),
\enskip i,j=1,2,\enskip i\not= j\$.

One should note that while \$U_a\$ maps
the range of \$\rho_1\$ onto that of
\$\rho_2\$, its adjoint \$U_a^{\dag}\$,
equalling its inverse \$U_a^{-1}\$,
maps the latter range onto the former.
Mathematicians would write, e. g.,
\$U_a\circ\rho_1^{1/2}\$ etc. \$'\circ
'\$ meaning "after" because the
operators do not act in one and the
same space.\\

{\bf Proof.} Any bipartite state vector
can be expanded in any complete
orthonormal first-subsystem basis, and
the (generalized) expansion
coefficients are the images of the
corresponding basis vectors by the
(antilinear) operator \$U_a\rho_1
^{1/2}\$, which are vectors in
\$\cH_2\$:
$$\ket{\Psi}_{12}=\sum_q\Big[\ket{q}_1
\Big(U_a\rho_1^{1/2}\ket{q}_1\Big)_2
\Big]$$ \cite{FHMV76}, \cite{FHZurek}
(section 2). The rest is standard
evaluation:
$$\tr\Big(\ket{\Psi}_{12}\bra{\Psi}_{12}
P_1P_2\Big)=\bra{\Psi}_{12}\Big(
\sum'_q\ket{q}_1\bra{q}_1P_2\Big)
\ket{\Psi}_{12}.$$ Substitution of the
expanded form of the state vector in
conjunction with (4) leads to the
claimed result.\hfill $\Box$\\

Though the correlations in an
\$N$-partite state \$\rho\$ are, in
general, an evasive entity, their
measure, showing how much of them there
is, is given by many authors in various
forms. The present author is partial to
the von Neumann entropy, and the {\it
correlation information} \$I_{12\dots
N}\$ as a measure of correlations that
follows from it.

Let a CD be given (cf Definition 1). On
account of the universal law of
subadditivity of entropy, one has the
definitions:
$$I_{12\dots N}\equiv
\Big(\sum_{i=1}^NS_i\Big)-S_{12\dots
N},\eqno{(5a)}$$ where \$I_{12\dots
N}\$ is the correlation information in
\$\rho\$, \$S_i\$ is the von Neumann
entropy of the \$i$-th particle and
\$S_{12\dots N}\$ is the von Neumann
entropy of the entire system. Further,
$$I_{C_k}\equiv \sum_{i\in
C_k}S_i-S_{C_k}\quad k=1,2,\dots
,n,\eqno{(5b)}$$ where \$"C_k"\$
denotes the \$k$-th cluster,
\$I_{C_k}\$ denotes its correlation
information, and \$S_{C_k}\$ is its
entropy; finally, $$I_{\sqcap}\equiv
\sum_{k=1}^nS_{C_k}-S_{12\dots N}
\eqno{(5c)}$$ is the among-the-clusters
correlation information.

The following theorem follows easily
(cf Theorem 1 in \cite{FH04}):
$$I_{12\dots
N}=\sum_{k=1}^nI_{C_k}\enskip+\enskip
I_{\sqcap}.\eqno{(5d)}$$ In words: The
total correlation information is the
sum of the within-the-cluster ones,
summed over all clusters, and the
among-the-cluster one.\\

\vspace{0.5cm}

\noindent

{\bf \noindent 5 . The uncorrelated
cluster decompositions}\\

{\bf Definition 3.} We call any cluster
decomposition (cf Definition 1) that
implies tensor factorization of the
state of the composite system into the
states (reduced density operators) of
the subsystems $$\rho
=\prod_{k=1}^{\otimes
n}\rho_k\eqno{(6)}$$ an {\it
uncorrelated cluster decomposition}
(UCD).\\

Now we formulate and prove a basic
consequence of the Wootters-Mermin
theorem, the {\it UCD theorem}.

{\bf Theorem 2.} One is dealing with an
uncorrelated cluster decomposition {\it
\IF every string} of subsystem events
\$P_1P_2\dots P_{n}\$ is {\it
uncorrelated}: $$\tr(\rho P_1\dots
P_{n})=\prod_{k=1}^{n}\tr(\rho_kP_k)
\eqno{(7)}$$ (cf (4)). Equivalently, if
there is a string of subsystem events
that 'sees' correlations, then and only
then one is not dealing with a UCD.\\

{\bf Proof.} {\it Necessity.} Assuming
the tensor factorization (6), one has
$$\tr\Big((\prod_{k=1}^{\otimes n}\rho_k)
(\prod_{k=1}^{\otimes
n}P_k)\Big)=\prod_{k=1}^n
\tr(\rho_kP_k).\eqno{(8)}$$

{\it Sufficiency.} Let (8) hold true
for all corresponding strings. Since,
according to the Wootters-Mermin
theorem, they determine a unique
composite-system state, it is obviously
\$\prod_{k=1}^{\otimes n}\rho_k\$.
\hfill $\Box$\\

{\bf Definition 4.} If one has two
cluster decompositions of the system
considered, one says that the latter
decomposition is a {\it coarsening} of
the former if each class in the latter
consists of one or several classes of
the former. One also says that the
former decomposition is a refinement of
the latter. The corresponding
adjectives are: 'coarser' and 'finer'.\\

If one cluster decomposition is a
coarsening of another, then the former
CD is, obviously, also a decomposition
into classes of the set of clusters of
the
latter, finer cluster decomposition.\\

Now we formulate the {\it FUCD
theorem}, the basic result of this
investigation.

{\bf Theorem 3. A)} For every
composite-system state \$\rho\$, there
exists, in a {\it unique way, a finest
uncorrelated cluster decomposition}, i.
e., one that implies (6) and that is
such that every other UCD is its
coarsening.

{\bf B)} One has an uncorrelated
cluster decomposition {\it if and only
if} it is a {\it coarsening}
of the finest cluster decomposition.\\

Before we prove the theorem, we need
some auxiliary insight.\\

{\bf Lemma 2.} Every cluster
decomposition of \$\{1,2,\dots ,N\}\$
{\it induces} a decomposition into
subclusters {\it in every subset} of
\$\{1,2,\dots ,N\}\$ by requiring that
two particles within the subset belong
to the same subcluster if they belong
to the same
cluster in \$\{1,2,\dots ,N\}$.\\

{\bf Proof} is obvious.\hfill $\Box$\\

{\bf Lemma 3.} Every uncorrelated
cluster decomposition induces (cf Lemma
2) in  every subset of \$\{1,2,\dots
,N\}\$ an uncorrelated subcluster
decomposition.\\

{\bf Proof.} Let an uncorrelated CD (cf
Definition 3) and a subset of
\$\{1,2,\dots ,N\}\$ be given. Let,
further, \$P_1P_2\dots P_n\$ be the
product of arbitrary projectors on the
induced subclusters in the subset. (If
the subcluster corresponding to an
index \$1\leq k\leq n\$ is an empty
set, then we put \$P_k\equiv I\$, the
identity operator in \$\cH_{12\dots
N}\$.) The string of events 'sees' the
following correlations (cf (4))
$$\Big|\tr\Big(\rho
(\prod_{k=1}^nP_k)\Big)-\prod_{k=1}^n
\tr\Big(\rho_kP_k\Big)\Big|,$$ where
\$\rho_k\$ is the state (reduced
density operator) of the \$k$-th
subcluster.

One has $$\tr\Big(\rho_kP_k\Big)=
\tr\Big(\rho_k'P_k\Big),\quad
k=1,2,\dots ,n,\eqno{(9)}$$ where
\$\rho_k'\$ is the state of the cluster
whose intersection with the given
subset the subcluster is, because one
obtains the lhs from the rhs by partial
tracing. (Note that \$P_k,\enskip
k=1,2,\dots ,n\$, are subcluster
events). Hence, the string 'sees' the
same correlations in the subclusters as
in the clusters, where, according to
the UCD theorem (Theorem 2), no
correlations are 'seen'. Thus,
according to the same theorem, the
subclusters are in
uncorrelated states.\hfill $\Box$\\

Next, we realize that {\it
continuation} of a UCD is a UCD. More
precisely, one has the following claim.

{\bf Lemma 4.} If each cluster in a
given uncorrelated cluster
decomposition is further decomposed in
an uncorrelated way into subclusters,
then the entire decomposition of
\$\{1,2,\dots ,N\}\$ into subclusters
is an uncorrelated cluster
decomposition.\\

{\bf Proof} becomes evident when one
substitutes each \$\rho_k\$ in \$\rho =
\prod_{k=1}^{\otimes n}\rho_k\$ (cf
(6)) by its corresponding tensor
factorized form.
\hfill $\Box$\\

{\bf Lemma 5.} The intersection of two
uncorrelated cluster decompositions is
a UCD.\\

{\bf Proof.} If a UCD is intersected
with another UCD, then, according to
Lemma 2, each cluster of the former is
decomposed into subclusters. Further,
according to Lemma 3, the
decompositions are uncorrelated.
Finally, according to Lemma 4, the
decomposition of \$\{1,2,\dots ,N\}\$
into the subclusters (intersections of
clusters) is a UCD. \hfill $\Box$\\

{\bf Lemma 6.} The intersection of any
number of uncorrelated cluster
decompositions is a UCD.\\

{\bf Proof.} Let us have \$L\$ UCD's,
\$2<L<\infty\$. We order them, in an
arbitrary but fixed way, into a
sequence. The intersection can be
obtained in a stepwise way by
intersecting the first UCD with the
second, the result of this with the
third etc. We apply Lemma 5 at each
step.\hfill $\Box$\\

We need just one more auxiliary
result.\\

{\bf Lemma 7.} Every coarsening (cf
Definition 4) of a UCD is a UCD.\\

{\bf Proof} becomes evident if one
makes an isomorphic transition from
\$\cH_1\otimes\cH_2\otimes\dots\otimes
\cH_N\$ to an \$N$-particle Hilbert
space ordered according to the two
given CD's: Consecutive groups of
particles correspond (in an arbitrary
way) to clusters of the first CD, and
groups of clusters according to the
coarsening are also consecutive. Then
it is seen that substitution of the
groups of states in (6), rewritten
according to the new ordering, by the
states corresponding to the coarsening
does not change the tensor-product
structure of the relation.\hfill $\Box$\\

{\bf Proof of the FUCD theorem.}

{\bf A)} According to Lemma 6, the
intersection of all UCD's is a UCD. It
is obvious that it is the finest of all
of them.

{\bf B)} It is obvious from claim A)
that every UCD must be a coarsening of
the finest one. Conversely, that every
coarsening of the finest UCD is a UCD
is a consequence of Lemma 7.\hfill
$\Box$\\

{\bf Definition 5.} We say that the
state of a subsystem is {\it
homogeneous} if it appears among the
tensor factors in the finest
uncorrelated decomposition (cf (6)).
Otherwise it is said to be
{\it heterogeneous}.\\

Next, we study states of homogeneous
clusters.\\

\vspace{0.5cm}

\noindent {\bf 6. Dynamical study.}\\

\indent  Influence of subsystem
measurement and unitary evolution on
the opposite subsystem in a bipartite
system is going to be investigated.\\

\vspace{0.5cm}

\noindent {\bf 6.A Ideal subsystem \m
.}\\

We begin with the simple, textbook case
of \m . We have a discrete
decomposition of the identity
\$I=\sum_{q=1}^QP_1^q\$, \$Q\$ finite
or infinite (\$\forall q,q':\enskip
P_1^qP_1^{q'}=\delta_{q,q'}P_1^q=
\delta_{q,q'}(P_1^q)^{\dag}\$).\\

Let us consider a bipartite system in a
state \$\rho_{12}\$. Let us, further,
consider an {\it ideal first-subsystem
\M } in the non-selective version (when
one deals with the entire ensemble)
ascertaining which of the events
\$P_1^q\$ will occur on each element of
the ensemble. When this \M is
performed, then the state (reduced
density operator) of subsystem \$2\$
does not change, but becomes decomposed
(distant state decomposition)
$$\rho_2=\sum_{q=1}^Q
\Big(\tr(\rho_{12}P_1^q)\Big)
\rho_2\Big\{\rho_{12},P_1^q\Big\},
\eqno{(10a)},$$ where the probabilities
\$\tr(\rho_{12}P_1^q)\$ are the
weights, and
\$\rho_2\Big\{\rho_{12},P_1^q\Big\}\$
are the admixed states defined as
follows
$$\forall q,
\enskip\tr(\rho_{12}P_1^q)>0:
\quad\rho_2\Big\{\rho_{12},P_1^q\Big\}
\equiv \tr_1(\rho_{12}P_1^q)\Big/
\tr(\rho_{12}P_1^q),\eqno{(10b)}
$$ where \$"\tr_1"\$ denotes the
partial trace over \$\cH_1$.\\

To prove claims (10a,b), remember that
ideal \M in its non-selective version,
by definition, gives a change of state
according to the L\"{u}ders formula
\cite{Lud}, \cite{FHMIN},
\cite{FHMIN2}:
$$\rho_{12}\quad\rightarrow\quad
\sum_{q=1}^QP_1^q\rho_{12}P_1^q.
\eqno{(11)}$$ The second-subsystem
state is obtained by tracing out
subsystem \$1\$ (and taking into
account that \$P_1^q\$ commutes under
the partial trace \$\tr_1\$ with the
other factor \$(\rho_{12}P_1^q)\$ and
that \$P_1^q\$ is idempotent). Thus one
obtains (10a) with (10b).\\

We call decomposition (10a) of the
state of subsystem \$2\$ 'distant'
because the measuring apparatus is
assumed to interact only with subsystem
\$1\$, and not at all with subsystem
\$2\$. Therefore, any influence of this
\M on subsystem \$2\$ is due
exclusively to the correlations between
subsystems \$1\$ and \$2\$. The term
'distant' should capture this
circumstance (in analogy with the case
when there is large spatial distance
between the subsystems).

Distant state decomposition (10a) is
just one of the countless
mathematically possible decompositions
of \$\rho_2\$. Its {\it physical
meaning} comes to the fore in the
selective aspect of the same \m .\\

When we consider the same \M in the
selective version, i. e., when the
state of individual systems, elements
in the ensemble, are the object of
description, then the composite system
undergoes the change
$$\rho_{12}\quad \rightarrow\quad
P_1^q\rho_{12}P_1^q
\Big/\tr(\rho_{12}P_1^q)\eqno{(12a)}$$
(unless \$P_1^q\$ is a probability-zero
event). As a consequence of this ideal
occurrence of \$P_1^q\$, the state of
the second subsystem is subject to the
(violent) change
$$\rho_2\quad\rightarrow\quad
\rho_2\Big\{\rho_{12},P_1^q\Big\}
\eqno{(12b)}$$ (cf(10b)). \\

If the bipartite state is uncorrelated,
i. e., if \$\rho_{12}=\rho_1
\otimes\rho_2\$, then $$\forall
q,\enskip\tr(\rho_{12}P_1^q)>0: \quad
\rho_2\{\rho_{12},P_1^q\}=\rho_2,
\eqno{(13)}$$ (as it is obvious from
(10b)), and the distant state
decomposition (10a) is trivial. In
other words, if two subsystems are
uncorrelated, ideal \M in one of them
cannot influence the other.

Thus, {\it the finest uncorrelated
decomposition discloses the boundaries
where the distant influence of
subsystem \M stops}.\\

There is a serious objection to the
importance of the conclusion just
reached. Namely, ideal \M is
overidealized; it is hard to make use
of it in the laboratory. More general
\m s are the realistic ones
\cite{FHpolar}. One wonders if the
boundaries of uncorrelatedness stop
also the influence of more realistic \m
s.\\

\vspace{0.5cm}

\noindent {\bf 6.B Realistic \m s.}\\

\indent In realistic \m s the change of
state is far more complicated than in
the ideal case \cite{FHpolar}. It is
therefore desirable to avoid it. We
make a plausible assumption, and then
resort to classical probability theory.

{\bf Assumption.} If we have a
bipartite state \$\rho_{12}\$ and a
coincidence of subsystem events
\$P_1^qP_2\$ (cf Subsection 6.A for the
notation) in an arbitrary \m , then the
coincidence probability {\it equals}
the product of the probability of the
event \$P_1^q\$ and the probability of
\$P_2\$ in the state of subsystem \$2\$
that comes about as a result of the
ideal occurrence of \$P_1^q\$.
Naturally, we assume that \$P_2\$
occurs immediately after \$P_1^q\$.
More precisely, we take the limes when
the time interval between these two
occurrences goes to zero.

The Assumption amounts to equating
coincidence with the conditional
probability formula in classical
probability theory. One must keep in
mind Mermin's warning \cite{Mermin1}
that one must be very cautious when
interpreting conditional probability in
\qm , because the condition, a
correlatum in Mermin's terms, does not
exist (except in the trivial case when
the conditional probability equals the
original one).

The classical formula is easily
understood in terms of relative
frequencies in the \m . Let \$M_{12}\$
be the frequency of the coincidences
\$P_1^qP_2\$, \$M\$ the total number of
occurrences in the \m , \$M_1\$ the
frequency of \$P_1^q\$, and, finally,
\$M_2\$ that of \$P_2\$. Then, as well
known, the coincidence probability is
the limes of \$M_{12}\Big/M\$ when
\$M\$ goes to infinity.

One can write
\$M_{12}\Big/M=(M_1\Big/M)(M_{12}\Big/
M_1)\$ (unless \$M_1=0\$). Taking the
limes separately of each factor on the
rhs, one comes to the mentioned
classical conditional probability
formula. Thus, the 'condition' is well
defined: we have in mind the cases when
\$P_1^q\$ occurs, then it is a given
correlatum, and it can serve the
purpose of a condition.

In the \QMl formalism the conditional
probability argument goes as follows.
$$\tr(\rho_{12}P_1^qP_2)=\Big(
\tr(\rho_{12}P_1^q)\Big)\Big[
\tr\Big(\rho_2\{\rho_{12}P_1^q\}P_2\Big)
\Big] \eqno{(14)}$$ (cf (10b)).

The argument presented establishes the
fact that the distantly prepared state
of subsystem \$2\$, when \$P_1^q\$
occurs in realistic \m , is the same as
in ideal \m . Hence also the important
conclusion that {\it boundaries of
uncorrelatedness stop measurement
influence} is valid in the
general case.\\

\vspace{0.5cm}

\noindent {\bf 6.C Subsystem
interaction has no distant influence}\\

The claim in the title of the
subsection is proved by the following
elementary argument.

Let \$\rho_{012}\$ be a tripartite
system such that
\$\rho_{12}\equiv\tr_0\rho_{012}\$ is a
bipartite system under investigation.
These entities apply to an initial
moment:
\$\rho_{012}\equiv\rho_{012}(t=0)\$.

We assume that subsystems \$0\$ and
\$1\$ interact in an arbitrary way, but
that subsystem \$(0+1)\$ does not
interact with the (hence distant)
subsystem \$2\$. We express this by the
tensor product of unitary (dynamical
evolution) operators $$\rho_{012}\quad
\rightarrow\quad\rho_{012}(t)\equiv
\Big(U_{01}(t)\otimes
U_2(t)\Big)\rho_{012}
\Big(U_{01}(t)^{\dag}\otimes
U_2(t)^{\dag}\Big).$$

Then we have
$$\rho_2(t)\equiv\tr_{01}
\Big(\rho_{012}(t)\Big)=
\tr_{01}\Big[\Big(U_{01}(t)\otimes
U_2(t)\Big)
\rho_{012}\Big(U_{01}(t)^{\dag}\otimes
U_2(t)^{\dag}\Big)\Big]=$$
$$U_2(t)\Big[
\tr_{01}\Big(U_{01}(t)\rho_{012}
U_{01}(t)^{\dag}\Big)\Big]U_2(t)^{\dag}=
U_2(t)\Big[\tr_{01}\Big(\rho_{012}
U_{01}(t)^{\dag}U_{01}(t)\Big)\Big]
U_2(t)^{\dag}=$$ $$U_2(t)
\Big(\tr_{01}(\rho_{012})\Big)
U_2(t)^{\dag}=
U_2(t)\rho_2U_2(t)^{\dag}.$$

This known fact can be put so that
whatever goes on dynamically with the
state of subsystem \$1\$ when no
dynamical influence is exerted on
subsystem \$2\$, it {\it does not
influence the latter via correlations
either}. This general claim is derived
after the \M influences of the
preceding two subsections because of
the striking {\it contradiction}.
Namely, one expects \M to be a special
case of dynamical evolution. (This is
the so-called paradox of the \Q theory
of \m .)

The way I see it, there are two basic
schools of thought in foundational \qm
. The first stipulates that unitary
dynamical evolution is not universal;
altered dynamics applies to \M
\cite{GRW}.

The second school of thought sticks to
exclusively unitary dynamics, but it
abandons the 'prejudice' of absolute
properties to which we are used from
classical physics, particularly from
special relativity theory. Jordan
\cite{Jordan}, in a skilful variation
of Hardy's approach to Bell's theorem
\cite{Hardy}, proves that the
assumption of local and real properties
- which is the same as absolute
properties - of the famous EPR argument
\cite{EPR} is in contradiction with \qm
.

To avoid the paradox, it takes some
kind of relative-state view in the
spirit of Everett \cite{Everett}, like,
e. g., the
relative-reality-of-unitarily-evolving-state
(RRUES) view as it was expounded in
recent \QMl discussions of the
delayed-choice erasure experiments of
Scully et al. \cite{FHScully1},
\cite{FHScully2}. We will resume this
point of view in the next section after
we present a beautiful EPR-type
entanglement case, introduced in the
after-Mermin investigations by Cabello.\\

\vspace{0.5cm}

\noindent {\bf 7. EPR-type entanglement}\\

Cabello \cite{Cabello59} suggested to
consider a quadri-partite purely-spin
state vector that is the tensor product
of two singlet states:
$$\ket{\Psi}_{1234}
\equiv\Big((1/2)^{1/2}(\ket{+}_1\ket{-}_2
-\ket{-}_1\ket{+}_2)\Big)\otimes
\Big((1/2)^{1/2}(\ket{+}_3\ket{-}_4
-\ket{-}_3\ket{+}_4)\Big).\eqno{(15)}$$
In view of the FUCD theorem, it is easy
to see that this state is written in
its finest uncorrelated decomposition
form (because the decomposition cannot
be continued), and that there is no
other non-trivial CD ((15) does not
have a non-trivial coarsening, cf the
FUCD theorem).

The subsystems \$(2+3)\$ and \$(1+4)\$,
which are bipartite in their turn, are
correlated in (15). We first change the
order of the tensor factors in (15) (by
isomorphism) from \$1234\$ to \$2314\$.
Then, as easily seen, (15) can be
rewritten in the (isomorphic) form as a
(maximally correlated) Schmidt
canonical decomposition \cite{FHZurek}
(section 2).
$$\ket{\Psi}_{2314}'=
(1/2)\bigg[\Big(\ket{-}_2\ket{+}_3\Big)\otimes
\Big(\ket{+}_1\ket{-}_4\Big)\enskip
+\enskip\Big(\ket{-}_2\ket{-}_3\Big)
\otimes\Big(-\ket{+}_1\ket{+}_4\Big)\enskip
+$$
$$\Big(\ket{+}_2\ket{+}_3\Big)\otimes
\Big(-\ket{-}_1\ket{-}_4\Big)\enskip
+\enskip\Big(\ket{+}_2
\ket{-}_3\Big)\otimes
\Big(\ket{-}_1\ket{+}_4\Big)
\bigg].\eqno{(16)}$$

Since the eigenvalue \$1/4\$ of the
reduced density operator \$\rho_{23}\$
(as well as that of \$\rho_{14}\$) has
fourfold degeneracy, expansion of
\$\ket{\Psi}_{2314}'\$ in any basis in
the four-dimensional space
\$\cH_2\otimes\cH_3\$ gives a canonical
Schmidt decomposition (cf
\cite{FHZurek}, section 2). The basis
in \$\cH_2\otimes\cH_3\$ in which
\$\Psi'_{2314}\$ is expanded in (16) is
uncorrelated. Cabello rightly suggests
\cite{Cabello59} to take an opposite
case, i.e., one with a basis of
maximally correlated state vectors. The
well-known Bell states
$$\ket{\psi^{\pm }}_{23}\equiv
(1/2)^{1/2}\Big(\ket{+}_2
\ket{-}_3\enskip\pm\enskip\ket{-}_2
\ket{+}_3\Big),\eqno{(17a)}$$
$$\ket{\phi^{\pm }}_{23}\equiv
(1/2)^{1/2}\Big(\ket{+}_2\ket{+}_3\enskip
\pm\enskip\ket{-}_2\ket{-}_3\Big),
\eqno{(17b)}$$ which also form an
orthonormal basis in
\$\cH_2\otimes\cH_3\$, are quite
suitable.

The evaluation of the 'partner' in each
term of a Schmidt canonical
decomposition is much facilitated by
the use of the (antiunitary) {\it
correlation operator} \$U_a\$ that is
uniquely implied by every bipartite
state vector. It is an {\it invariant}
entity for all Schmidt canonical
decompositions of a given bipartite
state vector, and it maps precisely the
first-tensor-factor basis vectors into
the second ones, into their 'partners'
in the decomposition \cite{FHZurek}
(Appendix A there).

Therefore, it is practical to start
with the Schmidt canonical
decomposition that is an expansion in
the uncorrelated basis as (16) is, read
\$U_a\$ in it, and then one can
immediately write down the 'partners'
in any other Schmidt canonical
decomposition. In this way one obtains:
$$\ket{\Psi}_{2314}'=(1/2)\bigg[
\ket{\psi^+}_{23}\otimes
\ket{\psi^+}_{14}\enskip+\enskip
\ket{\psi^-}_{23}\otimes
\Big(-\ket{\psi^-}_{14}\Big)\enskip+$$
$$\ket{\phi^+}_{23}\otimes
\Big(-\ket{\phi^+}_{14}\Big)\enskip+\enskip
\ket{\phi^-}_{23}\otimes\Big(-
\ket{\phi^-}_{14}\Big)\bigg].\eqno{(18)}$$

If the subsystems are spatially
sufficiently far away from each other
so that one can perform subsystem \m s,
which by definition must not
dynamically influence the opposite (or
distant) subsystem, then any Schmidt
canonical decomposition has an
important {\it physical meaning}. (In a
purely spin state as (15) is, one can
safely assume the feasibility of
subsystem \M because in the suppressed
spatial tensor-factor part of the state
the two particles can be far away from
each other.)

Let us take (16). If one performs on
the nearby subsystem \$(2+3)\$ a \M to
ascertain in which of the uncorrelated
states (first tensor factors in (16))
the subsystem is, then {\it ipso
facto}, by distant, i. e., by
interaction-free \M the distant
subsystem \$(1+4)\$ finds itself in the
uncorrelated 'partner' state.
(Schr\"{o}dinger \cite{Schroed35},
\cite{Schroed36} would say that the
distant subsystem is 'steered' into the
'partner' state.)

If we decide to measure on the
subsystem \$(2+3)\$ in which of the
(maximally correlated) Bell states
(17a,b) it is, then, a look at (18)
tells us that {\it ipso facto} one
finds out by distant \M in which of the
same Bell states the distant subsystem
\$(1+4)\$ is. This is EPR-type
disentanglement, the heart of the
famous EPR paradox \cite{EPR}.

Note that the two mentioned direct
subsystem \m s are mutually
incompatible, and usually they are
considered as alternative choices. But
the real random delayed-choice erasure
experiment of Kim et al. \cite{Kim} has
shown that it is possible to perform
the two mutually incompatible \m s in
one and the same experiment (cf also
the \QMl insight in the experiment
\cite{FHScully2}).

Also Seevinck mentions the above
distant \M of Bell states, but he views
it as entanglement swapping
(\cite{Seevinck}, Section 5, (i)).

EPR-type disentanglement is a striking
example of what the correlations can do
if a boundary appearing in a UCD does
not stop it (like in the bipartite
system \$(1+2)+(3+4)\$, cf (15), in
contrast to \$(2+3)+(1+4)\$, cf (16)).

Returning to the second school of
thought in foundational \qm , which
maintains the exclusiveness of unitary
evolution, mentioned in the preceding
section, we can repeat the Ithaca
mantra of Mermin: "The correlations,
not the correlata". In the above case,
both (16) and (18) simultaneously
really exist in the given quadripartite
state vector along with infinitely many
other (also incompatible) Schmidt
canonical decompositions. The 'partner'
subsystem states in each term express
aspects of the correlation, which is
best encapsulated in the correlation
operator \$U_a\$, which covers all
possible Schmidt canonical
decompositions. Measurements only add
new subsystems (the measuring
instruments) to make a more complex
multipartite state vector, but
essentially nothing is changed. (See
the \QMl insight in delayed-choice
erasure experiments in \cite{FHScully1}
and \cite{FHScully2}.)

The "not the correlata" part of
Mermin's mantra means, the way I
understand it, that the tensor factors
in the components of e. g. (16) or (18)
or any other concrete Schmidt canonical
decomposition, or rather the elementary
events that they define, cannot be
considered real in an absolute sense.
In the standard \QMl language , they
are potentialities. (This corresponds
to the more usual claim that
observables do not have definite values
in such cases.)

In the
relative-reality-of-unitarily-evolving-state
(RRUES) approach, which is in the
spirit of Everett \cite{Everett}, and
which seems to be required by the
exclusively unitary evolution, {\it the
correlations}, along with the particles
of which the systems are made up,
appear to be {\it the basic building
blocks of reality}.\\

\vspace{0.5cm}

\noindent {\bf 8. Correlational
isolation or being correlationally
closed}\\

{\bf Definition 6.} A state \$\rho\N\$
of a system of \$N\$ particles is {\it
correlationally isolated} (from its
environment) or {\it correlationally
closed} if, whenever \$K\$ particles
from the environment, \$1\leq K\$, are
joined to the \QMl description, in the
state of the enlarged system (of
\$N+K\$ particles) the original system
of \$N\$ particles is uncorrelated with
the \$K\$ added ones: $$\rho_{12\dots
N(N+1)\dots
(N+K)}=\rho\N\otimes\rho_{(N+1)\dots
(N+K)},\eqno{(19)}$$ where the factors
are the corresponding subsystem states
(reduced density operators). Otherwise
the state of the system is
correlationally open or unisolated.\\

We state and prove now a result {\it on
subsystem inheritance}.

{\bf Proposition.} If
\$\rho_{12}=\rho_1 \otimes\rho_2\$ is
an uncorrelated state of a bipartite
system, then also the state of each
subsystem of subsystem \$1\$ is
uncorrelated with the state of any
subsystem of subsystem
\$2\$.\\

{\bf Proof} follows immediately when
one takes the partial traces ( in the
above tensor product) over the
particles that do not belong to the
(smaller) subsystems considered.
\hfill $\Box\$\\

{\bf Corollary 2.} If the state of a
system is correlationally isolated from
its environment, then so is the state
of its every subsystem.\\

{\bf Proof.} The preceding proposition
immediately implies Corollary 2.\hfill
$\Box$\\

If a cluster state is homogeneous in
the state \$\rho\$ of the \$N$-particle
system and it is also correlationally
isolated from the environment of the
latter, irrespectively if so is also
\$\rho\$, then we say that the state of
the cluster is {\it absolutely
homogeneous}.

It should be noted that if one
considers the finest uncorrelated
decomposition of a correlationally open
system, each of the homogeneous
subsystems can be, independently of
each other, open or closed
correlationally. We have seen in the
preceding section what apparently
devastating influence the correlations
can transfer from the nearby subsystem
to the distant one.\\

\vspace{0.5cm}

\noindent {\bf 9. Comments on
Seevinck's and Cabello's articles}\\

Now we take a critical look from the
point of view of the CD, UCD, and FUCD
theorems of this article at some
mentioned important work.

To my knowledge, Jordan \cite{Jordan}
was the first to take a critical view
of the Ithaca program \cite{Mermin1}.
It was pointed out (in the last-but-one
passage of Section 6) that this author
made an important contribution to
understanding distant correlations. His
criticism of Mermin is similar to that
of Cabello, and the latter is more
clearly articulated. Therefore, I wont
discuss Jordan's critical attitude
separately.\\

\noindent {\bf 9.A Seevinck's article}\\

\noindent In \cite{Seevinck} the title
reads: "The Quantum World is not Built
up from Correlations". This claim
should be contrasted with the FUCD
theorem (Theorem 3) of this study,
which says that every state of a
several-particle system is the tensor
product of homogeneous disjoint cluster
states, i. e., of states that cannot be
further tensor factorized, and that
this is unique. Hence, the
composite-system state is built up from
homogeneous cluster states. There are
no correlations between these clusters,
but the cluster states contain
correlations.

In the Introduction of \cite{Seevinck}
the question of anticorrelation of spin
projections in the singlet state is
raised and it is asked "whether or not
we can think of this (anti-)
correlation as a real property of the
two-particle system independent of
measurement". Later on it is stated
that "in this letter we will
demonstrate that ... no such
interpretation is possible".

The singlet two-particle state, as any
pure state, is a tensor factor in the
state of any cluster containing the two
particles (it is absolutely homogeneous
in the terms of this study). There is
no reason why the mentioned
anti-correlation could not be viewed as
a piece of reality of nature. (We
return to this  below.)

To my mind, the central point in
Seevinck's article is his assumption of
"local realism". He expresses it for a
four-partite system that he views as a
bipartite system the subsystems of
which are, in turn, each bipartite. It
reads (cf his relation (6)):
$$P_{\hat A\hat B,\hat C\hat
D}(ab,cd|W_0)=P^I_{\hat A\hat B}
(ab|W_I)P^{II}_{\hat C\hat
D}(cd|W_{II}),\eqno{(20)}$$ where
\$W_0\$ is the state (density operator)
of the composite system, and
\$W_i,\enskip i=I,II\$ are the states
(reduced density operators) of the
subsystems. Further, \$\hat A,\hat
B,\hat C, \hat D\$ are observables
(Hermitian operators) of the four
finest subsystems, and \$a,b,c,d\$ are
their possible eigenvalues. The lhs
$$P_{\hat A\hat B,\hat C\hat
D}(ab,cd|W_0)=\tr\Big(W_0\hat A\hat B
\hat C\hat D\Big)$$ is the average
value of the product of the four
observables, and on the rhs we have the
product of corresponding averages in
the two (larger) subsystems.

Since Hermitian operators are linear
combinations of their spectral
projectors (if one confines oneself to
observables with finite spectra), it is
sufficient to restrict the four
observables to projectors.

Remembering the UCD theorem (Theorem
2), one can expect (20) to be valid if
the two larger subsystems are, as
clusters, uncorrelated, i. e., if
\$W_0=W_I\otimes W_{II}\$. Otherwise,
one would expect that one can choose
the four projectors so that (20) is not
valid.

If one finds the alleged "local
realism" condition (20) not valid, this
has nothing to do with "realism", only
with lack of tensor factorization (lack
of uncorrelatedness). It does have to
do with "localness" because Seevinck
uses this term as a synonym for
subsystem: "Local thus refers to being
confined to a subsystem of a larger
system, without requiring the subsystem
itself to be localized (it can thus
itself exist of spatially separated
parts)." (See the first footnote in
\cite{Seevinck}.) We saw in Sections 6
and 7 that the state of a subsystem is
'untouchable' in the sense of distant
manipulation only if it is uncorrelated
with the complementary cluster. Hence,
(20) could pass as a 'locality'
condition.

The author derives from (20) a
Bell-like inequality with the intention
to violate it. Violation of a necessary
condition (the inequality) of relation
(20) implies violation of (20) itself.
I confine my discussion to (20).

Let us return to the singlet state
mentioned at the beginning of this
section. If it is combined with another
bipartite pure or mixed state (it can
be combined only by tensor product)
into a four-partite system, it will
satisfy Seevinck's "local realism"
condition as obvious from the UCD
theorem (Theorem 2). Hence the
spin-projection anti-correlation in the
singlet state can be interpreted as
local and real.\\

The author takes a particular state
vector
$$\ket{\Psi }=(1/2)^{1/2}\Big(\ket{
\uparrow\downarrow\uparrow\downarrow }
-\ket{\downarrow\uparrow\downarrow
\uparrow }\Big)\eqno{(21)}$$ of a
four-partite system. Then he shows that
the Bell-like inequality, hence also
(20), is violated.

For the validity of (20) it is a
relevant question if (21) is
homogeneous or heterogeneous. We show
now that the former is true.

The tensor-product states \$\ket{
\uparrow\downarrow\uparrow\downarrow
}\$ and
\$\ket{\downarrow\uparrow\downarrow
\uparrow }\$ are quadri-orthogonal, i.
e., orthogonal in each of the four
factors. Hence, it is written in a
maximally entangled Schmidt canonical
form (\cite{FHZurek}, section 2) how
ever one makes the quadri-partite
system bipartite. It is thus seen that
\$\ket{\Psi }\$ of (21) is not a tensor
product of two two-partite subsystems.
No single-particle reduced density
operator is a ray projector. Hence,
(21) cannot contain a single-particle
tensor factor state vector.

We conclude in this way that the finest
uncorrelated cluster decomposition of
(21) is the trivial decomposition. No
wonder that (20) does not hold for it.

Now I make a short relevant deviation.
As it is well known, the numerous
Bell-like inequalities for hidden local
values of observables, which were
statistical, i. e.,  applied to
ensembles, were superseded by
equalities proving Bell's theorem
saying that \QM does not allow an
extrapolation with local hidden values
of all local observables for {\it
individual systems}. This came later.
But as early as in 1977 Stapp has
proved an interesting theorem saying
that the assumption that individual
systems have definite local values of
all local observables is in
contradiction with \QM (\cite{Stapp}).

Mermin has postulated (as part of the
Ithaca program \cite{Mermin1}) that
probabilities should have some physical
meaning for individual systems. Hence,
Seevinck's plausible Bell-like
inequality for correlations makes me
conjecture that

(i) if one were able to extend the
(vague) \Q correlations
idea to a
subquantum level, then

(ii) one might be able to derive a
Stapp-like theorem saying that
correlations in parts of individual
clusters that have a homogeneous state
in the state of a supersystem could be
distantly manipulated from other parts
of the cluster as in section 6. In this
sense I could agree with Seevinck that
individual-system correlations (only
their hypothetical subquantum
extension) in subclusters of clusters
with a homogeneous state lack "robust
reality". But this is a bit far fetched.\\

\vspace{0.5cm}

\noindent {\bf 9.B Cabello's
articles}\\

In \cite{Cabello59} the author presents
the EPR-type disentanglement described
in Section 7. He uses this example to
attack Mermin's Ithaca program arguing
as follows (p. 114, left column).

"Mermin's interpretation assumes {\it
physical locality}, defined as "[t]he
fact that the internal correlations of
a dynamically isolated system do not
depend on any interactions experienced
by other systems external to it.""
Cabello claims that the mentioned
EPR-type disentanglement refutes
Mermin's "physical locality". Somewhat
below this the author continues his
argument (having in mind (18)).

"By this violation of physical locality
I do not mean that the internal
correlations between particles \$1\$
and \$4\$ "change" after a spacelike
separated experiment (this does not
happen in the sense that no new
internal correlations are "created"
that were not "present" in the reduced
density matrix for the system \$1\$ and
\$4\$ before any interaction), but that
the type of internal correlations (and
therefore, according to Mermin, {\it
the reality}) of an individual isolated
system {\it can be chosen at
distance}." (Only the last italics are
due to the present author.)

Now I analyse Cabello's criticism
leaning on the present study and on the
concrete example of Cabello's mentioned
EPR-type disentanglement.

(i) As it was shown in subsection 6.C,
no unitary interaction of an external
system with subsystem \$(2+3)\$ can
influence the state \$\rho_{14}\$ of
the opposite cluster \$(1+4)\$. In this
sense Mermin's quoted 'physical
locality' is satisfied.

(ii) If one interprets the mentioned
EPR-type disentanglement in an
Everett-like relative-state way (cf
Sections 6 and 7), then all that
happens when the Bell-states \M is
performed on the subsystem \$(2+3)\$ is
that also the \M apparatus becomes
correlated with subsystem \$(1+4)\$ (cf
an analogous discussion in more detail
in \cite{FHScully1}). Then again
Mermin's point of view is unshaken.

(iii) If one interprets the distant
effects of the Bell-states \M on
\$(2+3)\$ assuming collapse (and
modification of the unitary dynamical
law), then Mermin's 'physical locality'
is refuted. In this sense Cabello's
criticism seems justified.\\

Further, Cabello claims (p.114, second
column): "A consistent interpretation
(he means the Ithaca program, FH) could
be developed by keeping correlations as
fundamental but avoiding to say that
they are local properties." In view of
the mentioned EPR-type correlations,
particularly item (iii) above, this
claim seems plausible. Though I, for my
part, would add to "fundamental" also
"real".

Cabello seems partial to the Copenhagen
interpretation of \QM and he says:
"Mermin's proposal can be seen as an
attempt to complete the Copenhagen
interpretation." In contrast to this
view, it seems to me that the Ithaca
program [1] can be understood as seeing
{\it reality} in the quantum state and
{\it in the correlations} it contains,
and not just being "a purely symbolic
procedure" according to words
of Bohr as Cabello quotes them.\\

It seems to me that Cabello in his
second paper \cite{Cabello60} assumes
the interpretation with collapse (cf
item (iii) above) and elaborates the
untenability of 'local' correlations.\\

\vspace{0.5cm}

\noindent {\bf 10. Conclusion}\\

\noindent It was stated in the
discussion of Seevinck's article that
it seems likely that individual-system
correlations in parts of clusters that
have a homogeneous state may lack
'robustness of reality' because they
can be distantly, i. e., without
interaction, 'changed' (in the collapse
approach). The point to note is that
this applies to {\it individual
systems}. How ever strongly we wish to
understand the physics of individual
systems, \QM is, unfortunately, the
physics of {\it ensembles} of \Q
systems. (As well known, even Einstein
had no quarrel with this. He only
claimed that physics of individual \Q
systems will require more than just the
\Q state. We do not know if he was
right or not.)

The present study along the lines of
Mermin's ideas treats correlations only
through probabilities, and these are
observed ensemblewise. Leaning on this
fact, one can claim that the reality of
correlations {\it is locally robust}.
In other words, the mentioned
manipulation of one subcluster by
another (as in Sections 6 and 7) is a
global effect, i. e., it takes place in
the supercluster (with a homogeneous
state). Locally there is no
manipulation.\\

Mermin in the first place and to the
largest extent, but also Jordan,
Cabello and, particularly, Seevinck
began a very intriguing investigation
into fundamental many-partite \qm . I
have joined this line of research with
the intention to help to lift the mist
shrouding the field. Time will tell to
what extent our efforts, jointly taken,
have achieved this. But I think that we
have at least turned the mist into
haze; and this, if it lasts, will not
be so difficult to disperse.\\

\vspace{0.5cm}

\noindent {\bf Appendix}

The aim is to characterize {\it density
matrices} in a finite, \$M$-dimensional
unitary space. We'll do it, following
Wootters \cite{Wootters}, representing
them in a linearly independent basis
consisting of projectors.\\

To begin with, we study the concept of
linear independence.

{\bf Lemma} The following three
properties of a set of \$Q\$ elements
\$\{P_q:q=1,2,\dots ,Q\}\$ of a
\$Q$-dimensional unitary space are {\it
equivalent}.

(i) If \$\sum_q\omega_qP_q=0\$, then
necessarily \$\omega_q=0,\enskip
q=1,2,\dots ,Q\$.

(ii) Let \$\{A_r:r=1,2,\dots ,Q\}\$ be
an orthonormal basis. The transition
matrix \$\alpha\$, the elements of
which appear in $$P_q=
\sum_{r=1}^Q\alpha_{qr}A_r,\enskip
q=1,2,\dots ,Q\eqno{(A.1)}$$ is
non-singular, or equivalently, its
determinant is non-zero.

(iii) The matrix
\$\{\tr\Big(P_qP_{q'}\Big)
:q,q'=1,2,\dots ,Q\}\$ is non-singular,
i. e., its determinant is non-zero.\\

{\bf Proof.}
\Big("(ii)"$\enskip\Leftrightarrow
\enskip$"(iii)"\Big):
\$\tr\Big(P_qP_{q'}\Big)=
\sum_r\sum_{r'}\alpha_{pr}\alpha_{q'r'}
\tr\Big(A_rA_{r'}\Big)\$. Since by
definition \$\tr\Big(A_rA_{r'}\Big)=
\delta_{rr'}\$, one obtains
\$\tr\Big(P_qP_{q'}\Big)=\sum_r
\alpha_{qr}\tilde\alpha_{rq'}\$, where
\$\tilde\alpha\$ is the transpose of
\$\alpha\$. This implies
\$det\Big[\tr\Big(P_qP_{q'}\Big)\Big]=
\Big(det[\alpha ]\Big)^2$. Hence,
\$det[lhs]\not= 0\enskip\Leftrightarrow
\enskip det[rhs]\not= 0$.\\

$\Big("(i)"\enskip\Leftrightarrow\enskip
"(ii)"\Big)$: Let us write down the
identity
$$\sum_q\omega_qP_q=\sum_q
\omega_q\sum_r\alpha_{qr}A_r=\sum_r
\Big(\sum_q\tilde\alpha_{rq}
\omega_q\Big)A_r,\eqno{(A.2)}$$ where
\$\{\omega_q:q=1,2,\dots ,Q\}\$ are any
numbers, and
\$\{\alpha_{qr}:q,r=1,2,\dots ,Q\}\$ is
the expansion matrix (A.1) (which need
not be non-singular at this stage). If
we put the identity (A.2) equal to
zero, then necessarily also
$$\sum_q\tilde\alpha_{rq}\omega_q=0,\quad
r=1,2,\dots ,Q\eqno{(A.3)}$$ is
satisfied due to the assumed
orthonormality of the basis
\$\{A_r:r=1,2,\dots ,Q\}$.

If we assume the validity of "(i)",
then it follows from relation (A.3)
that the matrix \$\tilde\alpha \$,
hence also \$\alpha\$ is non-singular,
i. e., that "(ii)" is valid.
Conversely, if requirement "(ii)" is
satisfied, then the matrix
\$\tilde\alpha\$ being non-singular,
relation (A.3) implies
\$\omega_q=0,\enskip q=1,2,\dots ,Q\$,
i.e., in view of the first expression
in (A.2), "(i)" is satisfied.\hfill
$\Box$

Any of the three requirements defines a
{\it linearly independent} sequence
\$\{P_q:q=1,2,\dots ,Q\}$.\\

We envisage the density operator
expanded \$\rho =\sum_{q=1}^{M^2}\chi_q
P_q\$ in \$M^2\$ linearly independent
projectors. (Now \$Q\equiv M^2\$.) On
the other hand, it is determined by
what one can measure, i. e., by the
probabilities \$\Big\{\tr
\Big(P_q\rho\Big):q=1,2,\dots
,M^2\Big\}\$. Substituting the expanded
form of \$\rho\$ in the probabilities,
one obtains $$\tr(P_q\rho )=\sum_{q'=1}
^{M^2}\chi_{q'}\tr(P_qP_{q'}), \quad
q=1,2,\dots ,M^2.\eqno{(A.4)}$$

On account of the non-singularity of
the matrix
\$\{\tr(P_qP_{q'}):q,q'=1,2,\dots
,M^2\} \$ (cf definition (iii) of
linear independence), it has an
inverse, and the expansion coefficients
\$\{\chi_q:q=1,2,\dots
,M^2\}\$ are uniquely determined.\\

The number of \$M^2\$, or rather of
\$(M^2-1)\$ probabilities if one takes
the identity operator as one of the
linearly independent projectors,
determines a Hermitian operator (as we
have argued in the text). One may
wonder how can Wootters be sure that
\$\rho\$ will be a density operator (a
non-negative operator), which is a
special case of Hermitian operators. My
conjecture is that Wootters leaned on
Gleason's theorem \cite{Gleason} in
this, which guarantees that one must be
dealing with a density operator.\\

\end{document}